\documentclass[
amsmath,
amssymb,
aps, 
prl,
twocolumn, 
groupedaddress,
nobalancelastpage,
floatfix]
{revtex4-2}
\usepackage{mathtools}
\usepackage[colorlinks]{hyperref}
\usepackage{lipsum}
\usepackage{bbold}
\usepackage{bm}
\usepackage{dsfont}
\usepackage{enumerate}
\usepackage{graphicx}
\usepackage{float}
\usepackage {xcolor}
\usepackage{cleveref}
\usepackage{amsmath}

\usepackage{lipsum}

\renewcommand{\H}{\hat{H}}
\newcommand{\A}{\hat{\vec{A}}}

\renewcommand{\vec}[1]{\bm{#1}}	

\DeclarePairedDelimiterX{\mean}[1]{\langle}{\rangle}{
	{#1}
}
\DeclarePairedDelimiterX{\abs}[1]{\lvert}{\rvert}{
	{#1}
}
\DeclarePairedDelimiterX{\norm}[1]{\lVert}{\rVert}{
	{#1}
}
\DeclarePairedDelimiterX{\bra}[1]{\langle}{\rvert}{#1}

\DeclarePairedDelimiterX{\ket}[1]{\lvert}{\rangle}{#1}

\DeclarePairedDelimiterX{\mel}[3]{\langle}{\rangle}{
	{#1}\delimsize\vert {#2}\delimsize\vert {#3}
}
\DeclarePairedDelimiterX{\inner}[2]{\langle}{\rangle}{
	{#1} \delimsize\vert{#2}
}
\DeclarePairedDelimiterX{\dyad}[2]{\lvert}{\vert}{
	{#1} \delimsize\rangle \delimsize \langle{#2}
}

\begin{document}
	\title{Excitonic Enhancement of Squeezed Light in Quantum-Optical High-Harmonic Generation From a Mott Insulator}
	
	\author{Christian Saugbjerg Lange} 
	\author{Thomas Hansen}
	\author{Lars Bojer Madsen}
	\affiliation{Department of Physics and Astronomy, Aarhus University, Ny Munkegade 120, DK-8000 Aarhus C, Denmark}
	\date{\today}
	\begin{abstract}
	The strong-field induced generation of nonclassical states of light is not only a subject of fundamental research but also has potential usage in quantum information science and  technology. The emerging field of strong-field quantum optics has developed ways of generating nonclassical states of light from the process of high-harmonic generation (HHG) at much wider frequency ranges and intensities than is typical for quantum optics. So far, however, no clear nonclassical signal at a distinct and unambiguous frequency has been predicted. Here, we study the  response from an exciton in a Mott-insulating system, using the extended Hubbard model. We find that the exciton plays a key role in the nonclassical response and generates  squeezing at the exciton energy. We relate this nonclassical response to the nonvanishing time correlations of the current operator in the system. Our work defines a direction for experimental work to search for squeezed light from HHG in a spectrally confined region defined by the  exciton energy. 
	\end{abstract}
	
	\maketitle
	
	
	For decades, a semiclassical description of strong-field processes with quantized electrons and classical electromagnetic fields has been successful. In recent years, a shift towards a fully quantized description including the quantized electromagnetic field has begun in an effort to merge the fields of strong-field physics and quantum optics. This emerging field of strong-field quantum optics enables the study of strong-field processes at a fundamental level, as it enables inquiries at the quantum level regarding, e.g., light-matter and photonic-mode entanglement. Such investigations bring new insights  and can lead to potential applications in quantum information at ultrafast timescales \cite{Lewenstein2024AttoAndQI}. 
	
	In quantum optics, one typically considers only a few photons and a few electronic states. In strong-field physics, the case is quite the opposite because a macroscopic number of photons is involved and the electronic Hilbert space involves many electronic states including the continuum. Consequently, new theoretical approaches to a quantum optical description of a macroscopic number of photons are needed. In recent years, various strong-field phenomena have been considered from a quantum optical perspective. For example, free electrons coupled to a quantized electromagnetic field can probe the photon statistics of the driving field \cite{DiGiulioFreeElectrons2019}, transfer optical coherence \cite{KfirFreeElectrons2021}, generate quantum light \cite{BenHayunKaminerFreeElectrons2021, DahanKaminerFreeElectrons2023}, have the quantum statistics of photons imprinted onto the electronic spectrum \cite{DahanKaminerFreeElectrons2021}, can be used for photonic state tomography \cite{GorlachKaminerPhotonTomographyFreeElectrons2024}, and even has applications in quantum information science \cite{RuimyKaminerFreeElectronQuantumOptics2025}. Additionally, other strong-field processes such as strong-field ionization have been considered with nonclassical driving fields for both single \cite{FangLiu2023} and double ionization \cite{Liu2025} of atoms, revealing how the nonclassical nature of the driver significantly influences the ionization process. 
	A key process in strong-field physics is high-order harmonic generation (HHG). Many works have surfaced studying HHG from a quantum optical perspective, revealing that the emitted light is indeed nonclassical from atoms \cite{Gorlach2020, Stammer2024a, Yi2024}, correlated solids \cite{Lange2024a, Lange2024b}, entangled systems \cite{Pizzi2023}, and can be engineered by subsequent measurements to generate an optical cat state \cite{Lewenstein2021, Lamprou2025}. Experimental works are confirming the nonclassical nature of light generated by HHG \cite{Gonoskov2016, Tsatrafyllis2017, Tsatrafyllis2019, Theidel2024, Lamprou2025}. Works have considered HHG with nonclassical driving fields, in particular bright-squeezed vacuum \cite{Manceau2019, Sharapova2020}, showing that the photon statistics of the driver drastically changes the emitted harmonic spectrum \cite{EvenTzur2023, Gorlach2023, EvenTzur2024, Gothelf2025}. Recently, experiments with nonclassical driving fields \cite{Rasputnyi2024, Lemieux2024, EvenTzur2025} have revealed how the nonclassical driver generates nonclassical states of light in HHG.

So far, when driven by a coherent state, all electronic systems considered have shown weak squeezing in the emitted light from HHG  at seemingly arbitrary frequencies with no clear, strong squeezing at a distinct and clearly identifiable wavelength \cite{Gorlach2020, Lange2024a, Lange2024b}. The lack of such a clear unambiguous signal at a well-defined frequency has hindered a clear identification of the relation between the electronic system and its nonclassical response and it also poses a challenge to future experimental work, as the broad frequency range with weak squeezing signal puts high demands on the experimental equipment. From semiclassical HHG studies, it has recently been found that excitons play a prominent role in the nonlinear response of the system \cite{Udono2022, MolineroExcitons2024, JensenExciton2024}. Here, we study the excitonic effects on the nonclassical response in HHG from Mott insulators. We find a relatively large degree of squeezing centered around the exciton energy. This response is drastically different from that reported earlier from an exciton-free Mott insulator both semiclassically \cite{Udono2022} and quantum optically \cite{Lange2024a, Lange2024b} and shows a clear mapping between the presence of the exciton and the nonclassical response. Mott excitons have been observed in, e.g., nickel compounds \cite{Ono2004, Ono2005} as well as Sr$_2$IrO$_4$ and Sr$_3$Ir$_2$O$_7$ \cite{Mehio2025}.
	
	\begin{figure}
		\centering
		\includegraphics[width=0.7\linewidth]{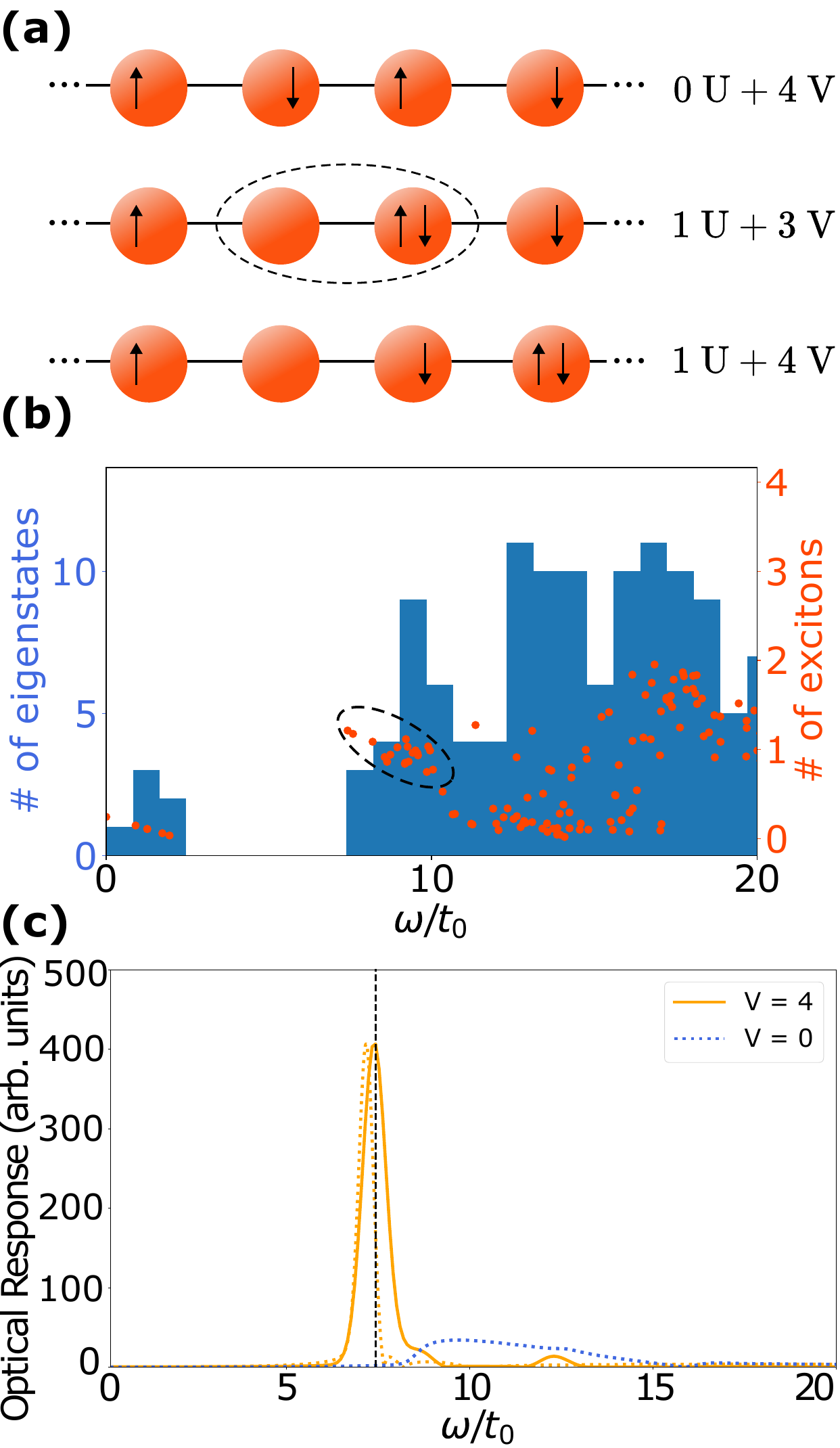}
		\caption{(a) Schematic of electron configurations and their potential energy obtained from Eq.~(\ref{eq:H_ee}) assuming periodic boundary conditions. $V$ effectively binds the doublon and holon giving rise to the exciton (dashed eclipse). (b) Energy histogram of the field-free eigenstates (blue) obtained from exact diagonalization for the low-energy regime of the system. The red dots show the expectation value of the number of excitons. (c) Linear optical response. A clear and distinct peak for the exciton system (orange) is seen, while the blue curve shows the system without the presence of excitons ($V=0$). The vertical dashed line indicates the exciton energy, while the dotted curves shows the linear response from an infinite system (see text) normalized to the linear response from the finite system (solid orange).}
		\label{fig:extendedhubbardthreefigures}
	\end{figure}
	

	We study the excitonic effects on the nonclassical response in HHG from a Mott insulator using the  extended Hubbard model (atomic units used throughout)
	\begin{equation}
		\H_{sc}(t) = \H_{hop}(t) + \H_{e-e} 
		\label{eq:H_SC}
	\end{equation}
	with kinetic energy from nearest-neighbor hopping modified by the driving field
	\begin{subequations}
		\begin{equation}
			\H_{hop}(t) = -t_0 \sum_{j, \mu} \big(e^{i aA_{cl}(t)} \hat{c}^\dagger_{j, \mu} \hat{c}_{j+1, \mu} + \text{h.c.} \big)
			\label{eq:H_hop}
		\end{equation}
	and potential energy from electron-electron repulsion
		\begin{equation}
		\H_{e-e}= U \sum_j \hat{n}_{j, \uparrow} \hat{n}_{j, \downarrow} +  V \sum_j \hat{n}_{j} \hat{n}_{j+1},
		\label{eq:H_ee}
		\end{equation}
	\end{subequations}
	where $\hat{c}_{j, \mu}$ ($\hat{c}^\dagger_{j, \mu}$) is the electronic annihilation (creation) operator on site $j$ with spin $\mu = (\uparrow, \downarrow)$, $\hat{n}_{j, \mu}  =  \hat{c}^\dagger_{j, \mu} \hat{c}_{j, \mu}$ is the electronic counting operator and $\hat{n}_{j} = \hat{n}_{j, \uparrow} + \hat{n}_{j, \downarrow}$. Following similar works, we set the nearest-neighbor hopping amplitude $t_0 = 0.0191$ a.u. and the lattice spacing $a = 7.5589$ a.u. \cite{Silva2018, Hansen2022a, Hansen2022b, Lange2024a, Lange2024, Lange2024b}.  $A_{cl}(t)$ is the classical driving field entering the Perierl's phase, and $U$ and $V$ are the parameters for the onsite and nearest-neighbor electron-electron repulsion, respectively. We consider a 1D cut along the linear polarization of $A_{cl}$, employ periodic boundary conditions, and assume half-filling with an equal number of both spin orientations. We limit ourselves to $L=8$ electrons, $U=12 t_0$, and $V=4 t_0$ which supports a spin-density-wave phase and the presence of excitons in the Mott insulator. For details on $U$ and $V$ and the related phases of the system, see Ref.~\cite{Jeckelmann2003}. The $V$ binds the empty sites (holons) to doubly occupied sites (doublons) giving rise to the Mott exciton. When $V=0$, there are no excitons in the system. The effect of $V \neq 0$ is illustrated in Fig.~\ref{fig:extendedhubbardthreefigures}(a), where three different electron configurations are shown with their respective potential energies [Eq. (\ref{eq:H_ee})] on the right (periodic boundary conditions). An electron hopping from the antiferromagnetic state which is present in the ground state of the system (top row) creates a doublon and a holon (middle row) next to each other. An additional electron hop separating the holon and doublon (bottom row) comes with an increase in potential energy from evaluating Eq.~(\ref{eq:H_ee}). In this sense, $V$ effectively binds the doublon and holon giving rise to the exciton. To further characterize the system, we obtain the field-free eigenstates of Eq.~(\ref{eq:H_SC}) from exact diagonalization and show the corresponding energies in Fig.~\ref{fig:extendedhubbardthreefigures}(b). Here, the lowest-lying states, including the ground state, are energetically separated by an energy gap making the system insulating. Further, we calculate the expectation value of the number of excitons in the field-free eigenstates, indicated by red dots in Fig.~\ref{fig:extendedhubbardthreefigures}(b). The lowest-lying states contain virtually no excitons, while the lowest energy states above the energy gap (marked by a dashed eclipse) contain approximately one exciton, which means that the energy gap from the ground state to the lowest-lying states across the gap is equal to the exciton energy. This is in line with the fact that the ground state is dominated by configurations with single-site occupations from which all first-order transitions result in the creation of an exciton and hence a substantial energy increase. The effect of the exciton is also clearly seen from the linear optical response obtained by applying a single-cycle, perturbative high-frequency pulse ($\omega \approx 10 t_0$). In fact, it is the transition energy between the ground state and exciton states that dominates the optical linear response as seen in Fig.~\ref{fig:extendedhubbardthreefigures}(c), where the single dominating peak for $V=4t_0$ corresponds exactly to the above-mentioned transition congruent with previous studies \cite{EsslerGebardJeckelmann2001, Jeckelmann2003, Udono2022}. The solid orange curve is obtained from exact diagonalization, while the dotted curves are obtained from infinite time-evolving block decimation calculations \cite{Vidal2007, Kjall2013} for an infinite system, showing that the strong exciton response (dashed orange) is still present in a large system and is not a finite-size effect. We note that without excitons in the system, $V=0$ (dashed blue), the linear optical response is a weaker continuum above the band gap without any clear peaks, showing that the response from free doublon-holon dynamics is drastically different from the exciton dynamics. A more detailed semiclassical analysis of the system in relation to HHG can be found in Ref.~\cite{Udono2022}.

 We now turn to a fully quantum optical description of the HHG process. After transforming away the coherent driving field, the Hamiltonian of the system is given as \cite{Gorlach2020, Lange2024a, Lange2024b}
\begin{equation}
	\H(t) = \H_{sc}(t) + \A\cdot \hat{\vec{j}}(t) + \H_F,
	\label{eq:H_full_QO}
\end{equation}
where $\H_{sc}(t)$ is the semiclassical Hamiltonian given in Eq.~(\ref{eq:H_SC}), $\A = \sum_{\vec{k}, \sigma} (g_0/\sqrt{\omega_k}) (\hat{a}_{\vec{k}, \sigma} + \text{h.c.})$ is the quantized vector potential in the dipole approximation with $g_0$ the coupling constant. The operator $\hat{a}_{\vec{k}, \sigma}$ annihilates a photon with momentum $\vec{k}$ and polarization $\sigma$ with $\hat{a}_{\vec{k}, \sigma}^\dagger$ being the corresponding creation operator, $\hat{\vec{j}}(t) = -ia t_0 \sum_{j, \mu} \big(e^{ia A_{cl}(t)} \hat{c}^\dagger_{j, \mu} \hat{c}_{j+1, \mu} - \text{h.c.} \big) \hat{\vec{x}}$ is the current operator taken along the direction of the chain, and $\H_F = \sum_{\vec{k}, \sigma} \omega_k \hat{a}_{\vec{k}, \sigma}^\dagger \hat{a}_{\vec{k}, \sigma}$ is the Hamiltonian of the free electromagnetic field. Inserting Eq.~(\ref{eq:H_full_QO}) into the time-dependent Schrödinger equation (TDSE), $i \partial_t \ket{\Psi(t)} = \H(t) \ket{\Psi(t)}$, and going to a rotating frame with respect to  $\H_{sc}(t)$ and $\H_F$, we obtain the equation of motion
\begin{equation}
	i \partial_t \ket{\chi^{(m)}_{\vec{k}, \sigma}(t)} = \A_Q^{(\vec{k}, \sigma)}(t) \cdot \sum_n \vec{j}_{m,n}(t) \ket{\chi^{(n)}_{\vec{k}, \sigma}(t)},
	\label{eq:chi_eom}
\end{equation}
where $\ket{\chi^{(m)}_{\vec{k}, \sigma}(t)} = \inner{\phi_m}{\Psi(t)}$ is the photonic state correlated to the $m$'th electronic eigenstate, $\ket{\phi_m}$, and $\vec{j}_{m,n}(t) = \bra{\phi_m} \hat{\vec{j}}_H(t) \ket{\phi_n}$ are current matrix elements with $\hat{\vec{j}}_H(t) = \mathcal{U}_{sc}^\dagger(t) \hat{\vec{j}}(t) \mathcal{U}_{sc}(t)$ a Heisenberg-type operator of the current operator with $\mathcal{U}_{sc}(t)$ the time-evolution operator for $\H_{sc}(t)$ of Eq.~(\ref{eq:H_SC}). In Eq.~(\ref{eq:chi_eom}), $\A_Q^{(\vec{k}, \sigma)}(t) = (g_0/\sqrt{\omega_k}) ( \hat{a}_{\vec{k}, \sigma}  e^{-i \omega_k t} + \text{h.c.})$ is the time-dependent operator for the vector potential acting on mode $(\vec{k}, \sigma)$, and the photonic modes are decoupled as in Refs.~\cite{Gorlach2020, Lange2024a, Lange2024b}   in order to be able to solve the dynamics numerically. We use a coherent driving pulse $A_{cl}(t) = (F_0/\omega_L) e^{-(t-t_c)^2/2\sigma_L^2} \sin[\omega_L(t- t_c)]$ with $F_0 = 0.0025$ a.u. corresponding to peak intensity of $2.2 \times 10^{11} $ W/cm$^2$ and $\omega_L = t_0 /2 = 0.00955$ a.u. $=396$ THz the carrier angular frequency of the pulse such that the exciton energy in Fig.~\ref{fig:extendedhubbardthreefigures}(c) is $\omega \approx 15 \omega_L$. The parameter $t_c = 5 T_L$ is the central peak time of the pulse, $\sigma_L = T_c$ is the pulse width where $T_L = 2\pi/\omega_L$, and we set $g_0 = 4 \times 10^{-8}$ a.u. \cite{Gorlach2020, Lewenstein2021, Lange2024a, Lange2024b}. We obtain $\vec{j}_{m,n}(t)$ from a semiclassical TDSE integration using the Arnoldi-Lancoz algorithm \cite{Park1986, Smyth1998, Guan2007, Frapiccini2014} with the dimension of the Krylov subspace set to $6$ \cite{Lange2024a, Lange2024b}. Once all currents are obtained, Eq.~(\ref{eq:chi_eom}) is solved by fourth-order Runge-Kutta integration by expanding the photonic state in a Fock basis truncated at maximum $50$ photons. All numerical results have been checked for convergence.


Before discussing the results, we consider a newly derived approximative expression for the photonic state, the so-called Markov-state approximation, first presented in Ref.~\cite{Stammer2024a} and numerically verified in Ref.~\cite{Lange2024b}. It is obtained by considering only current matrix elements involving the initial state, then performing a Markov approximation, and finally neglecting higher-order terms in $g_0$. The photonic state in the MSA is given as
\begin{equation}
	\ket{\chi^{(i)}_{\vec{k}, \sigma}(t) } = \hat{D}[\beta_{\vec{k}, \sigma}(t)] e^{-\langle \hat{W}_{\vec{k}, \sigma}^2(t) \rangle_{el}} \ket{0},
	\label{eq:chi_MSA_def}
\end{equation}
where $ \hat{D}[\beta_{\vec{k}, \sigma}(t)]= \exp[\beta_{\vec{k}, \sigma}(t) \hat{a}_{\vec{k}, \sigma}^\dagger - \beta_{\vec{k}, \sigma}^*(t) \hat{a}_{\vec{k}, \sigma}]$ is the displacement operator with  $\beta_{\vec{k}, \sigma}(t) = - i (g_0 /\sqrt{\omega_k}) \int^t dt' \langle \hat{\vec{j}}(t') \rangle \cdot \hat{\vec{e}}_\sigma^* e^{i \omega_k t'}$ and 
\begin{equation}
	\langle \hat{W}^2_{\vec{k}, \sigma}(t) \rangle_{el} = \int^t dt' \int^t dt'' \hat{A}_Q^{(\vec{k}, \sigma)}(t')\hat{A}_Q^{(\vec{k}, \sigma)}(t'') C(t', t''),
	\label{eq:W2_def}
\end{equation}
with $C(t',t'') = \langle \hat{\vec{j}}_H(t')\hat{\vec{j}}_H(t'') \rangle - \langle \hat{\vec{j}}_H(t) \rangle \langle \hat{\vec{j}}_H(t'') \rangle$ the time-correlation function of the current operator, $\hat{\vec{j}}_H(t)$. The subscript on the left-hand side of Eq.~(\ref{eq:W2_def}) denotes an expectation value only over the electronic degrees of freedom such that $\langle \hat{W}^2_{\vec{k}, \sigma}(t) \rangle_{el}$ is an operator in the photonic Hilbert space. Equation (\ref{eq:chi_MSA_def}) shows that if $C(t',t '') = 0$, i.e., if the currents are completely uncorrelated in time, the photonic state is a coherent state with no squeezing. On the other hand, if $C(t',t'') \neq 0$, Eq.~(\ref{eq:W2_def}) shows that the MSA contains quadratic photonic operators which generate squeezing \cite{Gerry2004, Cohen_Tannoudji1998_atomphoton, Scully1997}. The MSA description thus shows that the nonclassical response of the system is due to the time correlations of the induced current. 

We consider two observables in the HHG simulation: the spectrum and the squeezing. In the quantum optical formulation, the spectrum is given as \cite{Gorlach2020, Lange2024a}
\begin{equation}
	S(\omega) = \dfrac{\omega^3}{g_0^2 (2\pi)^2 c^3} \sum_\sigma \langle \hat{a}_{\vec{k}, \sigma}^\dagger \hat{a}_{\vec{k}, \sigma} \rangle,
	\label{eq:spectrum_general}
\end{equation}
where $\omega$ is the frequency of the mode $(\vec{k}, \sigma)$ and $c$ is the speed of light in vacuum. The degree of squeezing is given by the expression \cite{Scully1997, Braunstein2005}
\begin{equation}
	\eta_{\vec{k}, \sigma} = - 10 \log_{10}  \big\{4  \underset{\theta \in [0, \pi)}{\text{min}} [\Delta \hat{X}_{\vec{k}, \sigma}(\theta)]^2] \big\},
	\label{eq:squeezing_general}
\end{equation}
where the minimization is of the variance of the quadrature operator $\hat{X}_{\vec{k}, \sigma}(\theta) = (\hat{a}_{\vec{k}, \sigma} e^{-i\theta} + \text{h.c.})/2$. A nonvanishing squeezing, $\eta_{\vec{k}, \sigma} > 0$ shows the presence of nonclassical states of light \cite{Gerry2004}. 

\begin{figure}
	\centering
	\includegraphics[width=1\linewidth]{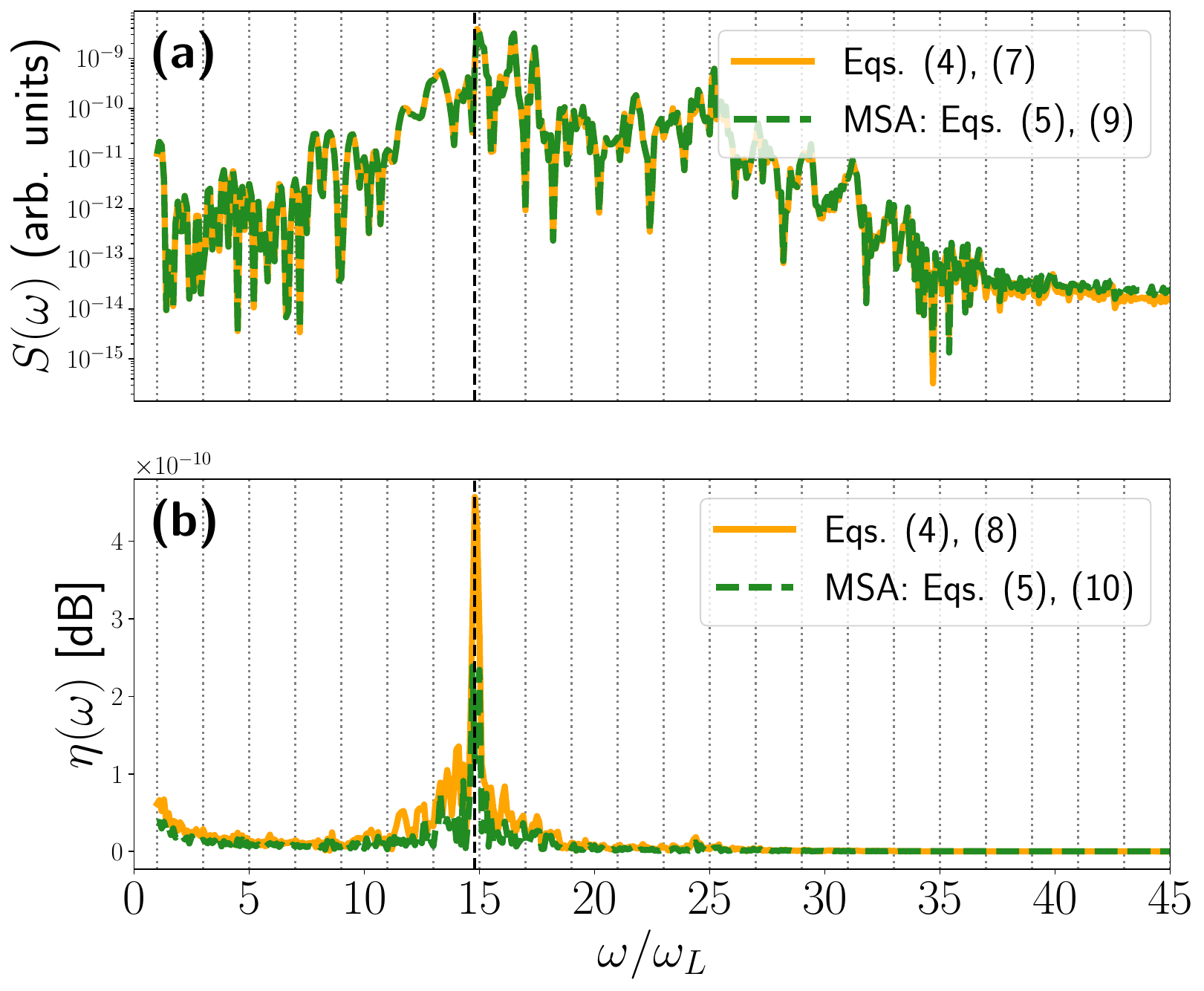}
	\caption{(a) Spectra obtained from Eqs. (\ref{eq:chi_eom}) and (\ref{eq:spectrum_general}) (orange) and the MSA expressions in Eqs. (\ref{eq:chi_MSA_def}) and (\ref{eq:spectrum_MSA}) (dashed green). The spectra peak at the exciton energy (vertical dashed line at $\omega \approx 15 \omega_L$)  showing that the exciton plays a significant role in the nonlinear response. (b) Squeezing from Eqs. (\ref{eq:chi_eom}) and (\ref{eq:squeezing_general}) (orange) and the MSA expressions from Eqs. (\ref{eq:chi_MSA_def}) and (\ref{eq:squeezing_MSA}) (dashed green). A clear peak in the signal is present at the exciton energy showing that the exciton is prominent in the nonclassical response of the system. The MSA captures the spectrum in (a) exactly and the squeezing to a good degree in (b). The vertical dotted lines at odd harmonics are added to guide the eye.}
	\label{fig:methodsalldatanormgaussianfulldatal8u12v4f1}
\end{figure}


Figure \ref{fig:methodsalldatanormgaussianfulldatal8u12v4f1} shows the results obtained from numerical integration of Eq.~(\ref{eq:chi_eom}) (solid orange) and from the MSA [Eq.~(\ref{eq:chi_MSA_def})] (dashed green). Figure \ref{fig:methodsalldatanormgaussianfulldatal8u12v4f1}(a) shows that the spectrum [Eq.~(\ref{eq:spectrum_general})] peaks at the exciton energy ($\omega \approx 15 \omega_L$), indicated by the vertical dashed line, showing that the exciton is highly involved in the nonlinear response \cite{Udono2022}. The spectrum is in general irregular with no clear peaks at odd harmonics even though the Hamiltonian in Eq.~(\ref{eq:H_SC}) is inversion symmetric. This irregularity is not a consequence of the quantum optical description of the HHG process but rather due to the population of multiple Floquet states during the non-adiabatic turn-on of the short pulse \cite{Lange2024}. As seen in Fig.~\ref{fig:methodsalldatanormgaussianfulldatal8u12v4f1}(a), the MSA calculation of the spectrum exactly matches the full calculation. The spectrum in the MSA is given as \cite{Lange2024b}
\begin{align}
	S^{\text{(MSA)}}(\omega) 	=  \dfrac{\omega^2}{(2\pi)^2 c^3} \sum_\sigma  \lvert \hat{\vec{e}}_\sigma^* \cdot \langle \hat{\vec{j}}(\omega)\rangle \rvert^2
	\label{eq:spectrum_MSA}
\end{align}
which is proportional to the square of the Fourier transform of the current and is hence similar to the semiclassical expression for the spectrum. In other words, the quantum optical description of HHG does not to leading order in the coupling constant, $g_0$, predict deviations from the classical spectrum \cite{Lange2024a, Lange2024b}.

The degree of squeezing calculated from Eq.~(\ref{eq:squeezing_general}) and shown in Fig.~\ref{fig:methodsalldatanormgaussianfulldatal8u12v4f1}(b) (orange line), shows a distinct peak at the exciton energy highlighting that the exciton dominantes the nonclassical response of the system. To analyze the origin of this squeezing, we use the expression from the MSA also illustrated in Fig.~\ref{fig:methodsalldatanormgaussianfulldatal8u12v4f1}(b)(green dashed line), which shows good agreement with the more exact calculation. In the MSA, the expression for the degree of squeezing is given as \cite{Lange2024b}
\begin{equation}
	\eta_{\vec{k}, \sigma}^{\text{(MSA)}} =  - 10 \log_{10}  \bigg\{\underset{\theta \in [0, \pi)}{\text{min}} [1 - 2 B_{\vec{k}, \sigma} \cos(2 \theta - \phi)] \bigg\},
	\label{eq:squeezing_MSA}
\end{equation}
where $\phi$ is a phase related to $B_{\vec{k}, \sigma} = (g_0^2/\omega) \lvert \int_{-\infty}^\infty dt' \int_{-\infty}^\infty dt'' \exp[i \omega(t'+t'')] C(t',t'') \rvert$. Equation (\ref{eq:squeezing_MSA}) shows that the degree of squeezing is determined by $B_{\vec{k}, \sigma}$ which is related to the time-correlation function $C(t',t'')$. Interestingly, a similar time-correlation function was considered in the Heisenberg picture in early work on HHG but not in the context of the nonclassical observables \cite{Sundaram1990}.  For $U\neq 0$ and $V\neq0$, the electrons are highly correlated, and  Eq.~(\ref{eq:squeezing_MSA}) shows that this correlation generates the squeezing, in accordance with the results presented in Fig.~\ref{fig:methodsalldatanormgaussianfulldatal8u12v4f1}(b) where it is clearly seen that the presence of an exciton is central to the nonclassical response of  the system. The reported degree of squeezing is low compared to  studies  assuming coherent participation of thousands of atoms \cite{Gorlach2020}. However, the present results are obtained with $L=8$ electrons and hence qualitatively show that the exciton is principal to the nonclassical response of the system. We believe the squeezing will be much larger for a macroscopic system or with an enhanced coupling strength, $g_0$. Our findings also encourage experimental work: If the linear response of such an excitonic system is measured, we predict that the light emitted from HHG at the exciton energy will have a relatively large degree of squeezing. This puts fewer demands on the experimental equipment as only a small frequency window has to be considered.


In summary, we have considered the nonclassical response of an exciton in a Mott-insulating system. This response has previously been studied semiclassically showing that the exciton plays a central role in the HHG spectrum \cite{Udono2022}. Here, we have gone beyond the semiclassical analysis and considered the full quantum optical description of the HHG process, and found that the exciton is also central in the nonclassical response as a relatively large degree of squeezing was found at the exciton energy. We interpreted this finding in the MSA and found that this nonclassical response is due to a nonvanishing time correlation of the current, which shows that the exciton is a key ingredient in the degree of correlation in the system. Previous studies have not shown a clear nonclassical response at a distinct and unambiguous energy, and as such, our findings motivate further work to investigate the squeezing of the light emitted from HHG in Mott insulators  at the exciton energy. 


This work was supported by the Independent Research Fund Denmark (Grant No. 1026-00040B). 
	
\bibliography{bib_this}

\end{document}